\documentclass[12pt]{article}
\usepackage{enumerate}
\usepackage{amsfonts}
\usepackage{amsmath}
\usepackage{amssymb}
\usepackage{amsthm}

\def\D{\mathcal{D}}

\def\H{\mathcal{H}}
\def\K{\mathcal{K}}

\def\S{\mathfrak{S}}
\def\F{\mathfrak{F}}
\def\C{\mathfrak{C}}
\def\T{\mathfrak{T}}
\def\B{\mathfrak{B}}

\newcommand{\rank}{\mathrm{rank}}

\newcommand{\id}{\mathrm{Id}}
\newcommand{\Tr}{\mathrm{Tr}}

\newcommand{\shs}{\hspace{1pt}}

\newcounter{defin}  \newcounter{lemma}  \newcounter{theorem}
\newcounter{property} \newcounter{corol}  \newcounter{remark} \newcounter{example}

\newenvironment{lemma}{\par\refstepcounter{lemma}
     \textbf{Lemma \thelemma.} }{\rm\par}
\newenvironment{theorem}{\par\refstepcounter{theorem}
     \textbf{Theorem \thetheorem.}\ }{\rm\par}
\newenvironment{property}{\par\refstepcounter{property}
     \textbf{Proposition \theproperty.}\ }{\rm\par}
\newenvironment{corollary}{\par\refstepcounter{corol}
     \textbf{Corollary \thecorol.} }{\rm\par}
\newenvironment{definition}{\par\refstepcounter{defin}
     \textbf{Definition \thedefin.}\ }{\rm\par}
\newenvironment{remark}{\par\refstepcounter{remark}
     \textbf{Remark \theremark.}}{\rm\par}

\begin{document}

\title{Energy-constrained diamond norms and quantum dynamical semigroups}

\author{M.E.~Shirokov, A.S. Holevo \\
\\
Steklov Mathematical Institute, Moscow, Russia}
\date{}
\maketitle

\vspace{-10pt}

\begin{abstract}
In the developing theory of infinite-dimensional quantum channels the relevance of the energy-constrained diamond norms
was recently corroborated both from  physical and information-theoretic points of view.
In this paper we study necessary and sufficient conditions for  differentiability with respect to these norms of the strongly continuous semigroups of quantum channels (quantum dynamical semigroups).
We show that these conditions can be expressed in terms of the generator of the semigroup.
We also analyze conditions for representation of a strongly continuous semigroup of quantum channels as an exponential series converging w.r.t.
the energy-constrained diamond norm. Examples of semigroups having such a representation are presented.
\end{abstract}

\section{Introduction}

In the developing theory of infinite-dimensional quantum channels the relevance of the energy-constrained diamond norms
was recently corroborated both from  physical and information-theoretic points of view \cite{Datta,Pir,SCT,W-EBN}. The choice of an appropriate metric on the set of quantum channels is important for analysis of characteristics of these channels and for the study of quantum dynamical semigroups. When dealing with finite-dimensional quantum channels it is natural to use the metric induced by the norm of complete boundedness (usually called ``the diamond norm'' \cite{Kit,Watrous,Wilde}). This metric  can be considered  as a  measure of distinguishability of quantum  channels via quantum measurements \cite[Ch.9]{Wilde}. However the topology generated  by  the diamond norm  is generally too strong for description of physically interesting perturbations of infinite-dimensional quantum channels \cite{W-EBN}. Mathematically, the inadequacy of the diamond norm metric in infinite dimensions can be seen from the Kretschmann-Schlingemann-Werner theorem  which says that closeness of channels in this metric means closeness of the Stinespring isometries of these channels in the operator norm \cite{Kr&W}. So, the diamond norm metric can not properly reflect "deformations" of the Stinespring isometry of a channel in the weaker topologies on the set of isometric operators, in particular, in the strong operator topology.

More adequate is the topology of the strong (pointwise) convergence generated by the
family of seminorms $\Phi\mapsto \|\Phi(\rho)\|_1$, where $\rho$ runs over all input states of the channel. The strong convergence of a sequence $\{\Phi_n\}$ of quantum channels to a channel $\Phi_0$ means that
$$
\lim_{n\rightarrow \infty}\Phi_n(\rho)=\Phi_0(\rho)\quad \textrm{for any state}\quad\rho.
$$
The strong convergence topology naturally appears in the study of strongly continuous quantum dynamical semigroups (QDS) -- the semigroups $\{\Phi_t\}_{t\geq 0}$ of quantum channels satisfying
$$
\lim_{t\rightarrow 0^+}\Phi_t(\rho)=\rho\quad \textrm{for any state}\quad\rho.
$$
These semigroups play a central role in the theory of open quantum systems, where they are used as a basic dynamical model for irreversible evolution \cite{Davies, GKS, H-SSQT, Lind}.

It was observed recently that the strong convergence topology  on the set of infinite-dimensional quantum channels is generated by the energy-constrained
diamond norm (ECD norm) introduced independently in \cite{SCT,W-EBN} (a slightly different version of ECD norm was used in \cite{Pir}).
These norms appeared to be a useful tool for quantitative continuity analysis of the basic capacities of energy-constrained infinite-dimensional channels \cite{SCT,W-EBN}. In the recent work \cite{Datta} it is shown that the ECD norms can be used effectively  in the study of strongly continuous QDS. In particular, the ECD norms allow to obtain sharp estimates for the rate of convergence for strongly continuous QDS, which give the new lower bounds on the minimal time needed for a quantum system to evolve from one quantum state to another (quantum speed limits \cite{SL,qsl}).

In fact, any strongly continuous QDS is  continuous w.r.t. the ECD norm induced by a positive operator with  discrete spectrum of finite multiplicity  \cite[Proposition 3]{SCT}. However, since the ECD norm is not an operator norm (in the sense of the  Banach space theory), one can not apply directly the Banach space theory of norm continuous semigroups to study the strongly continuous QDS, although one can expect that certain facts from this advanced theory can be used in our more general situation.

One of the differences of the ECD norm continuous semigroups as compared to the operator norm continuous semigroups is that the ECD norm continuity \emph{does not imply} the differentiability w.r.t. to this norm, i.e. the existence of the ECD norm bounded generator which provides, in particular, the exponential representation of the semigroup. The aim of this paper is to obtain necessary and sufficient conditions for  differentiability of the strongly continuous QDS with respect to the ECD norm and to analyze the conditions for a decomposition into an exponential series converging w.r.t. the ECD norm.

\section{Preliminaries}

\subsection{Basic notations}

Let $\mathcal{H}$ be a separable infinite-dimensional Hilbert space, $\mathfrak{B}(\mathcal{H})$
-- the algebra of all bounded operators on $\mathcal{H}$ with the operator norm $\|\!\cdot\!\|$ and $\mathfrak{T}(\mathcal{H})$ --
the Banach space of all trace-class operators on $\mathcal{H}$ with the trace norm $\|\!\cdot\!\|_1$ (the Schatten class of order 1) \cite{B&R,R&S}. Let
$\mathfrak{T}_{+}(\mathcal{H})$ be the cone of positive operators in
$\mathfrak{T}(\mathcal{H})$. Trace-class operators will be usually denoted by the Greek letters $\rho $, $\sigma $, $\omega $, ...
The closed convex subsets
$$
\T_{+,1}(\H)=\{\rho\in \T_+(\H)\,|\, \Tr\rho\leq 1\}\quad \textrm{and} \quad \S(\H)=\{\rho\in \T_+(\H)\,|\, \Tr\rho=1\}
$$
of the cone $\T_+(\H)$ are complete separable metric spaces with the metric defined by the trace norm.
Operators in $\S(\H)$ are called density operators or \emph{states} \cite{B&R,H-SCI}. Extreme points of $\S(\H)$ are 1-rank projectors called \emph{pure states}.

Denote by $I_{\H}$ the unit operator on a Hilbert space
$\mathcal{H}$ and by $\id_{\mathcal{\H}}$ the identity
transformation of the Banach space $\mathfrak{T}(\mathcal{H})$.

If quantum systems $A$ and $B$ are described by the Hilbert spaces  $\H_A$ and $\H_B$ then the composite system $AB$ is described by the tensor product of these spaces $\H_{AB}\doteq\H_A\otimes\H_B$. A state in $\S(\H_{AB})$ is denoted by $\rho_{AB}$, its partial states $\rho_A=\Tr_B\rho_{AB}$ and $\rho_B=\Tr_A\rho_{AB}$ (here and in what follows $\Tr_A$ denotes $\Tr_{\H_A}$, etc.)
\smallskip

We will consider  unbounded densely defined positive operators on $\H$ having discrete spectrum of finite multiplicity.
Following \cite{W-EBN} we will call such operators \emph{discrete}. In Dirac's notations any discrete operator ${G}$ can be represented as
\begin{equation}\label{H-rep}
G=\sum_{k=0}^{+\infty} E_k |\tau_k\rangle\langle\tau_k|
\end{equation}
with the domain $\mathcal{D}(G)=\{ \varphi\in\H\,|\,\sum_{k=0}^{+\infty} E^2_k|\langle\tau_k|\varphi\rangle|^2<+\infty\}$, where
$\left\{\tau_k\right\}_{k=0}^{+\infty}$ is the orthonormal
basis of eigenvectors of ${G}$ corresponding to the nondecreasing sequence $\left\{\smash{E_k}\right\}_{k=0}^{+\infty}$ of eigenvalues
tending to $+\infty$. \smallskip

A linear transformation $\Phi$ of the Banach space  $\T(\H)$ will be called
\emph{superoperator}. A superoperator $\Phi$ is called
\emph{Hermitian preserving} if $\Phi(\rho^*)=[\Phi(\rho)]^*$ for any $\rho$ in $\T(\H)$ \cite{Watrous}.
A superoperator $\Phi$ is called \emph{quantum channel} if it is trace preserving and completely positive \cite{H-SCI,Watrous, Wilde}.
A one parameter semigroup $\Phi_t$ of quantum channels such that
$$
\lim_{t\rightarrow0^+}\Phi_t(\rho)=\rho,\quad \forall \rho\in\T(\H),
$$
in the trace norm is called \emph{(strongly continuous) quantum dynamical semigroup}.

\subsection{Energy-constrained diamond  norms}

The norm of complete boundedness of a linear transformation of the algebra $\B(\H_A)$ (cf. \cite{Paul}) induces (by duality) the
"diamond norm"
\begin{equation}\label{d-n-def}
\|\Phi\|_{\diamond}\doteq\sup_{\rho\in\T(\H_{AR}),\|\rho\|_1\leq 1}\|\Phi\otimes \id_R(\rho)\|_1
\end{equation}
on the set of all superoperators on  $\T(\H_A)$, where $\H_R$ is a separable Hilbert space and $\H_{AR}=\H_{A}\otimes \H_{R}$ \cite{Kit}. If $\Phi$ is a Hermitian preserving  superoperator
then the supremum in (\ref{d-n-def}) can be taken over the set $\S(\H_{AR})$ \cite[Ch.3]{Watrous}.
\smallskip

The diamond norm is widely used in the quantum information theory, but generally the convergence induced by this norm is too strong for description of
physical perturbations of infinite-dimensional quantum channels: there exist quantum channels with close physical parameters such that the diamond norm distance between them is  equal to the maximal value $2$ \cite{W-EBN}. The reason of this inconsistency is pointed out briefly in the Introduction. By taking it into account the energy-constrained  diamond norms were introduced independently in \cite{SCT,W-EBN}.\footnote{Slightly different energy-constrained diamond norm is used in \cite{Pir}.}

\smallskip

Let ${G}$ be a positive  operator on $\H_A$ with a dense domain $\mathcal{D}({G})$ such that
\begin{equation}\label{G-cond}
\inf\left\{\shs\langle\varphi|{G}|\varphi\rangle\,|\,\varphi\in\mathcal{D}({G}),\|\varphi\|=1\shs\right\}=0.
\end{equation}
For any positive trace class operator $\rho$ the value of $\,\Tr \rho G$ (finite or infinite) is defined as $\,\sup_n \Tr \rho P_n G$, where $P_n$ is the spectral projector of $G$ corresponding to the interval $[0,n]$. If $\rho=\sum_k|\varphi_k\rangle\langle\varphi_k|$ then $\,\Tr \rho G=\sum_k\|\sqrt{G}\varphi_k\|^2$, where we assume that $\|\sqrt{G}\varphi_k\|=+\infty$ if $\varphi_k$ does not lie in $\mathcal{D}(\sqrt{G})$.

In applications $G$ is usually the Hamiltonian (energy observable) of the quantum system $A$. So, we will call the quantity
$\Tr \rho G$ the energy of a state $\rho$.\footnote{Sometimes, it is reasonable to assume that $G$ is some power of the Hamiltonian \cite{Datta}.}

Let $E>0$. The energy-constrained diamond  norm (ECD norm in what follows) of a  Hermitian-preserving  superoperator  $\Phi$ on $\T(\H_{A})$ is defined as
\begin{equation}\label{ECDN-d}
  \|\Phi\|^G_{\diamond,E}\doteq\sup_{\rho\in\S(\H_{AR}),\Tr \rho_AG\leq E}\|\Phi\otimes \id_R(\rho)\|_1,
\end{equation}
where $R$ is an infinite-dimensional quantum system (since all separable Hilbert spaces are isomorphic, this definition does not depend on $R$).  It is shown in \cite{W-EBN} that the nonnegative non-decreasing
function $E\mapsto\|\Phi\|^G_{\diamond,E}$ is concave on $[0,+\infty)$ for any given $\Phi$ and hence
$$
\|\Phi\|^{G}_{\diamond,E_1}\leq \|\Phi\|^{G}_{\diamond,E_2}\leq (E_2/E_1)\|\Phi\|^{G}_{\diamond,E_1}\quad\textrm{ for any } E_2>E_1>0.
$$
Thus, for given operator ${G}$ all the norms (\ref{ECDN-d}) are equivalent on the set  of all Hermitian preserving superoperators on $\T(\H_A)$.\smallskip

The quantity $\|\Phi\|^{G}_{\diamond,E}$  can be defined by formula (\ref{ECDN-d}) (as a positive number or $+\infty$) for any unbounded superoperator  $\Phi$ on $\T(\H_A)$ provided that the superoperator $\Phi\otimes \id_R$  is well defined on the set of states $\rho$ in $\S(\H_{AR})$ such that  $\Tr \rho_A G<+\infty$ (here $R$ is an infinite-dimensional quantum system). We will denote the set of all such superoperators by $\F_G(\H_A)$.
The arguments in \cite{W-EBN} showing concavity of the function $E\mapsto\|\Phi\|^G_{\diamond,E}$ remain valid for any $\Phi$ in $\F_G(\H_A)$. \smallskip

We will use the following simple observations.\smallskip

\begin{lemma}\label{ecdn-def} \emph{Let $\Phi$ be a superoperator in $\F_G(\H_A)$ and $E>0$.}\smallskip

A) \emph{The supremum in definition (\ref{ECDN-d}) can be taken over all operators in $\T_{+,1}(\H_{AR})$ satisfying the condition $\,\Tr \rho_A G\leq E$, i.e.}
\begin{equation}\label{ECDN-d+}
  \|\Phi\|^G_{\diamond,E}=\sup_{\rho\in\T_{+,1}(\H_{AR}),\Tr \rho_AG\leq E}\|\Phi\otimes \id_R(\rho)\|_1.
\end{equation}

B) \emph{Let $\,G_R$ be a positive operator on $\H_R$ unitarily equivalent to the operator $G$. Then
\begin{equation}\label{ECDN-d++}
\|\Phi\|_{\diamond,E}^G=\sup_{\rho\in\widehat{\S}_{G,G_R,E}}\|\Phi\otimes\id_R(\rho)\|_1,
\end{equation}
where
$\widehat{\S}_{G,G_R,E}\doteq\{\rho\in\S(\H_{AR})\,|\,\Tr \rho_A G\leq E,\, \Tr\rho_R  G_R\leq E,\, \rank\shs\rho=1\}$.}\smallskip
\end{lemma}\smallskip

\emph{Proof.} A) Since $\S(\H_{AR})\subset\T_{+,1}(\H_{AR})$, it suffices to show that $"\geq"$ holds in (\ref{ECDN-d+}).
Let $\rho$ be an operator in $\T_{+,1}(\H_{AR})$ such that $\,\Tr \rho_AG\leq E$ and $r=\Tr\rho$. Then $\hat{\rho}\doteq r^{-1}\rho$ is a state such that
$\,\Tr \hat{\rho}_AG\leq E/r$. So, by using concavity of the function $E\mapsto\|\Phi\|^G_{\diamond,E}$ on $\mathbb{R}_+$ and Lemma \ref{WL} below we obtain
$$
\|\Phi\otimes \id_R(\rho)\|_1=r\|\Phi\otimes \id_R(\hat{\rho})\|_1\leq r\|\Phi\|^G_{\diamond,E/r}\leq \|\Phi\|^G_{\diamond,E}.
$$

B) It suffices to show that $"\leq"$ holds in (\ref{ECDN-d++}). Since the system $R$ in definition (\ref{ECDN-d}) is assumed arbitrary and any mixed state in $\S(\H_{AR})$ can be considered as a partial state of some pure state in $\S(\H_{AR'})$, where $R'$ is an extension of $R$ \cite{H-SCI}, the supremum in (\ref{ECDN-d}) can be taken over all pure states $\rho$ in $\S(\H_{AR})$ satisfying the condition $\,\Tr \rho_A G\leq E$. Since for any such pure state $\rho$ the partial states $\rho_A$ and $\rho_B$ have the same nonzero spectrum,  by applying local partial isometry transformation of the system $R$ this state can be transformed into a state $\rho'$ belonging to the set
$\widehat{\S}_{G,G_R,E}$. It suffices to note that $\,\|\Phi\otimes\id_R(\rho)\|_1=\|\Phi\otimes\id_R(\rho')\|_1$.  $\square$ \smallskip

\begin{lemma}\label{WL} \cite{W-CB} \emph{If $f$ is a concave nonnegative function on $[0,+\infty)$ then for any positive $x< y$ and any $z\geq0$ the inequality $\,xf(z/x)\leq yf(z/y)\,$ holds.}
\end{lemma}\smallskip

The convergence on the set of quantum channels generated by any of the ECD norms implies  the strong  convergence:
\begin{equation}\label{c-rel}
 \lim_{n\rightarrow\infty} \|\Phi_n-\Phi_0\|^G_{\diamond,E}=0\quad \Rightarrow\quad\lim_{n\rightarrow\infty}\Phi_n(\rho)=\Phi_0(\rho)\,\textup{ for all }\rho\in\T(\H_A).
\end{equation}
If ${G}$ is a discrete  operator (\ref{H-rep}) then $"\Leftrightarrow"$ holds in (\ref{c-rel}) \cite[Proposition 3]{SCT}.
\smallskip

We will use the following \smallskip

\begin{lemma}\label{S-cont} \emph{If $\Phi$ is a superoperator in $\F_G(\H_A)$ such that  $\|\Phi\|^G_{\diamond,E}=o(\sqrt{E})$ as $E\rightarrow+\infty$
then for any separable Hilbert space $\H_R$ and any $E>0$ the superoperator $\,\Phi\otimes\id_R$ is uniformly continuous on the set of pure states $\rho$ in $\S(\H_{AR})$ such that $\Tr \rho_A G\leq E$. Quantitatively,
\begin{equation}\label{phi-cb}
\|\Phi\otimes \id_{R}(\rho-\sigma)\|_1\leq 2\varepsilon \|\Phi\|_{\diamond,2E/\varepsilon^2}^G
\end{equation}
for any pure states $\rho$ and $\sigma$ in $\S(\H_{AR})$ such that $\,\frac{1}{2}\|\rho-\sigma\|_1\leq\varepsilon$ and $\Tr \rho_A G,\Tr \sigma_A G\leq E$.}
\end{lemma}\smallskip

\emph{Proof.} The assumption of the lemma  implies that the r.h.s. of the inequality (\ref{phi-cb}) tends to zero as $\varepsilon\rightarrow0$. So, it suffices to prove this inequality.

Let $\rho$ and $\sigma$ be pure states in $\S(\H_{AR})$ such that $\Tr \rho_A G,\Tr \sigma_A G\leq E$ and $\frac{1}{2}\|\rho-\sigma\|_1\leq\varepsilon$. One can find pure states $\varrho$ and $\varsigma$ in some system $R'$ such that
$\varepsilon=\frac{1}{2}\|\hat{\rho}-\hat{\sigma}\|_1$, where $\hat{\rho}=\rho\otimes\varrho$ and $\hat{\sigma}=\sigma\otimes\varsigma$.
Then  $\alpha_{-}=\varepsilon^{-1}[\hat{\rho}-\hat{\sigma}]_{-}$ and $\alpha_{+}=\varepsilon^{-1}[\hat{\rho}-\hat{\sigma}]_{+}$ are pure states such that $\,\Tr[\alpha_{\pm}]_AG\leq 2E/\varepsilon^2$ \cite{AFM}. Since $\hat{\rho}-\hat{\sigma}=\varepsilon(\alpha_{+}-\alpha_{-})$, we have
$$
\begin{array}{rl}
\|\Phi\otimes \id_{R}(\rho-\sigma)\|_1\!\!\! & \leq\;
\|\Phi\otimes \id_{RR'}(\hat{\rho}-\hat{\sigma})\|_1=\,\varepsilon\|\Phi\otimes \id_{RR'}(\alpha_{+}-\alpha_{-})\|_1\\\\ & \leq\; \varepsilon\|\Phi\otimes \id_{RR'}(\alpha_{+})\|_1+\varepsilon\|\Phi\otimes \id_{RR'}(\alpha_{-})\|_1\leq \,2\varepsilon \|\Phi\|_{\diamond,2E/\varepsilon^2}^G.
\end{array}
$$
The first and second inequalities follow from the properties of the trace norm (non-increasing under partial trace and the triangle inequality), the third one -- from the definition of the ECD norm. $\square$

\subsection{Operator E-norms}

Let ${G}$ be a positive  operator on $\H$ with a dense domain $\mathcal{D}({G})$ satisfying the condition (\ref{G-cond}) and $E>0$.
The corresponding operator E-norm on $\B(\H)$ is defined as
\begin{equation}\label{ec-on}
 \|A\|^{G}_E\doteq \sup_{\substack{\rho\in\mathfrak{S}(\mathcal{H}):
\Tr \rho G\leq E}}\sqrt{\Tr A\rho A^*},\quad A\in \B(\H),
\end{equation}
where the supremum is over all states $\rho$ in $\S(\H)$ such that  $\Tr \rho G\leq E$. These norms are studied in detail in \cite{ECN}, where different applications of these norms are described.

One of the basic properties of the family of E-norms is the concavity of the nondecreasing function $E\mapsto\left[\|A\|^{G}_E\right]^2$ on $\mathbb{R}_+$ implying that
\begin{equation}\label{E-n-eq}
\|A\|^{G}_{E_1}\leq \|A\|^{G}_{E_2}\leq \sqrt{E_2/E_1}\|A\|^{G}_{E_1}\quad\textrm{ for any } E_2>E_1>0.
\end{equation}
Hence for given operator ${G}$ all the norms $\|\!\cdot\!\|^{G}_{E}$, $E>0$, are equivalent on $\B(\H)$.

Different  operators ${G}$ induce the E-norms generating different topologies  on $\B(\H)$. If $G$ is a discrete unbounded operator (\ref{H-rep}) then for any $E>0$ the norm $\|\!\cdot\!\|^{G}_{E}$ generates the strong operator topology on bounded subsets of $\B(\H)$ \cite[Proposition 2]{ECN}.

If $A$ is an unbounded linear operator with domain containing the set $\D(\sqrt{G})$  then its E-norm $\|A\|_E^G$ can be defined (as a nonnegative number or $+\infty$) by the same formula (\ref{ec-on}), where the supremum is taken over all \emph{finite rank} states $\rho$ satisfying the inequality $\Tr \rho G\leq E$
and $A\rho A^*$  is replaced by
\begin{equation}\label{ab-d}
  \sum_i|\alpha_i\rangle\langle\alpha_i|,\quad\quad |\alpha_i\rangle=A|\varphi_i\rangle,
\end{equation}
where $\rho=\sum_i |\varphi_i\rangle\langle\varphi_i|$ is a finite decomposition of $\rho$.\footnote{By using Schrodinger's mixture theorem (see \cite[Ch.8]{B&Z}) it is easy to show that the operator in (\ref{ab-d}) does not depend on this decomposition of $\rho$.}

This extension of the E-norm to unbounded operators is closely related to the notion of $\sqrt{G}$-relatively bounded operators  \cite{Kato,BS}. An operator $A$ is called relatively bounded w.r.t. the operator $\sqrt{G}$ (briefly, $\sqrt{G}$-bounded) if
$\mathcal{D}(\sqrt{G})\subseteq \D(A)$ and
\begin{equation}\label{rb-rel}
\|A\varphi\|^2\leq a^2\|\varphi\|^2+b^2\|\sqrt{G}\varphi\|^2,\quad \forall \varphi\in\mathcal{D}(\sqrt{G}),
\end{equation}
for some  nonnegative numbers $a$ and $b$. The $\sqrt{G}$-bound of $A$ (denoted by $b_{\sqrt{G}}(A)$ in what follows) is defined as
the infimum of the values $b$ for which (\ref{rb-rel}) holds with some $a$. If the $\sqrt{G}$-bound is equal to zero then $A$ is called $\sqrt{G}$-infinitesimal operator (infinitesimally bounded w.r.t. $\sqrt{G}$).

It is easy to show that $E\mapsto\|A\|_E^G$ is a finite function on $\mathbb{R}_+$ if and only if $A$ is a $\sqrt{G}$-bounded operator. Moreover, it is proved in \cite{RBO} that this
function coincides with the greatest lower bound of the functions $E\mapsto \sqrt{a^2+b^2 E}$ over all pairs $(a,b)$ for which (\ref{rb-rel}) holds
and that the $\sqrt{G}$-bound can be expressed as follows
\begin{equation}\label{G-bound}
  b_{\sqrt{G}}(A)=\inf_{E>0}\|A\|^{G}_{E}/\sqrt{E}=\lim_{E\rightarrow+\infty}\|A\|^{G}_{E}/\sqrt{E}.
\end{equation}

The set of all $\sqrt{G}$-bounded operators equipped with the norm $\|.\|_E^G$ is a Banach space denoted by $\B_G(\H)$ in \cite{ECN} (we identify operators coinciding on the set $\D(\sqrt{G})$). The set $\B(\H)$ is naturally embedded into $\B_{G}(\H)$ as a linear subspace, its closure $\B^0_{G}(\H)$ (i.e. the completion
of $\B(\H)$ w.r.t. the norm $\|.\|^{G}_E$) coincides with the set of all $\sqrt{G}$-infinitesimal operators \cite{ECN,RBO}.\smallskip

We will use the following lemma proved in \cite{SPM}.\smallskip

\begin{lemma}\label{qsl} \emph{Let $A$ and $B$ be any $\sqrt{G}$-infinitesimal operators  and}
\begin{equation}\label{s-1}
  \mathfrak{C}_{{G},E}\doteq \{\rho\in\T_{+}(\H)\,|\,\Tr\rho\leq 1,\, \Tr \rho G\leq E\shs\},\quad E>0.
\end{equation}

A) \emph{For any $\rho$ in $\mathfrak{C}_{{G},E}$ the operator $A\rho B^*\in\T(\H)$ is correctly defined by the formula
\begin{equation}\label{ab-d+}
  A\rho B^*\doteq\sum_i|A\varphi_i\rangle\langle B \varphi_i|,
\end{equation}
where $\rho=\sum_i |\varphi_i\rangle\langle\varphi_i|$ is any decomposition of $\rho$ into 1-rank operators, and}
\begin{equation}\label{star-in}
|\Tr A\rho B^*|\leq \|A\|^G_{E}\|B\|^G_{E}.
\end{equation}

B) \emph{The function $\rho\mapsto A\rho B^*$ is affine and uniformly continuous on the set $\,\C_{G,E}$ for any $E>0$.  Quantitatively,
\begin{equation}\label{ab-cb}
\|A\rho B^*- A\sigma B^*\|_1\leq \|A\|_E^G f_B(E,\varepsilon)+\|B\|_E^G f_A(E,\varepsilon)
\end{equation}
for any $\,\rho$ and $\sigma$ in $\C_{{G},E}$ such that $\|\rho-\sigma\|_1\leq\varepsilon$, where
$f_X(E, \varepsilon)=\sqrt{\varepsilon}\|X\|^{G}_{4E/\varepsilon}$ is a function vanishing as $\,\varepsilon\rightarrow 0^+$ for any $\sqrt{G}$-infinitesimal operator $X$.}\smallskip
\end{lemma}

The function $f_X(E, \varepsilon)$ tends to zero as $\,\varepsilon\rightarrow 0^+$, since formula (\ref{G-bound}) implies that $\|X\|^{G}_E=o(\sqrt{E})$ as $E\rightarrow+\infty$ for any $\sqrt{G}$-infinitesimal operator $X$.\smallskip

\begin{remark}\label{qsl-r} If $A$ and $B$ are $\sqrt{G}$-bounded  operators then inequality (\ref{star-in}) holds for any finite rank operator $\rho$ in $\mathfrak{C}_{{G},E}$ (provided that $A\rho B^*$ is defined by formula (\ref{ab-d+})).  If, in addition, $A$ and $B$ are closable operators then inequality (\ref{star-in}) holds for any $\rho$ in $\mathfrak{C}_{{G},E}$ \cite{ECN}.
\end{remark}\smallskip

We will also use the following lemmas.\smallskip

\begin{lemma}\label{+I} \cite{ECN} \emph{If $\,\K$ is any separable Hilbert space then  $\,\|A\otimes I_{\K}\|^{G\otimes I_{\K}}_E=\|A\|^G_E$.}
\end{lemma}\smallskip

\begin{lemma}\label{kln}\emph{Let  $A$ be a self-adjoint operator on $\H$. If the operator $A$ is $\sqrt{G}$-bounded then the operator $A^{p}$ is $\sqrt{G}$-infinitesimal for any $p\in[0,1)$ and $\,\|A^p\|^{G}_E\leq \left[\|A\|^G_E\right]^p$.}
\end{lemma}\smallskip

\emph{Proof.} By using the spectral representation of $A^2$, concavity of the function $x^p$ and the Jensen's inequality
we obtain $\Tr \rho A^{2p}\leq [\Tr \rho A^2]^{p}$ for any finite rank state $\rho$ with finite energy $\Tr \rho G$. So, $\,\|A^p\|^G_E\leq \left[\|A\|^G_E\right]^p$.  By using (\ref{G-bound}) and concavity of the function $E\mapsto\left[\|A\|^G_E\right]^2$ it is easy to show that $A^{p}$ is a $\sqrt{G}$-infinitesimal operator. $\square$\smallskip

\begin{lemma}\label{Gb}\emph{Let $A$ be a self-adjoint operator with domain containing the set $D(\sqrt{G})$. If $\Tr \rho A^2$ is finite for any state $\rho$ with finite energy $\Tr\rho G$ then the operator $A$ is\break $\sqrt{G}$-bounded.}
\end{lemma}\smallskip

\emph{Proof.} Suppose that the operator $A$ is not $\sqrt{G}$-bounded. Then $\,\|A\|^G_E=+\infty$ for any given $E>0$. Hence, there is a sequence $\{\rho_n\}$ of states
in $\S(\H)$ such that $\Tr\rho_n G\leq E$ and  $\Tr\rho_n A^2\geq 2^n$. Consider the state $\rho_*=\sum_{n=1}^{+\infty}2^{-n}\rho_n$. By  lower semicontinuity and convexity of the function $\rho\mapsto \Tr\rho G$ we have $\Tr\rho_* G\leq E$, while by using concavity and nonnegativity of the function $\rho\mapsto \Tr\rho A^2$ it is easy to show that $\Tr\rho_* A^2\geq \sum_{n=1}^{+\infty}2^{-n}\Tr\rho_n A^2=+\infty$. $\square$\smallskip

\section{Continuity and differentiability of quantum dynamical semigroups w.r.t. the energy-constrained diamond norm}

Let $\Phi_t$ be a (strongly continuous) quantum dynamical semigroup on  $\T(\H_{A})$. We will explore analytical properties of this
semigroup w.r.t. the metric induced by the ECD norm (\ref{ECDN-d}) assuming that $G$ is a discrete  operator (\ref{H-rep}) with $E_0=0$. \smallskip

The generator $S$ of a semigroup $\Phi_t$ is the superoperator
\begin{equation}\label{Gen-d-rel}
 S: \rho\mapsto \lim_{t\rightarrow0^{+}} t^{-1}(\Phi_t(\rho)-\rho)
\end{equation}
defined on the set $\D(S)$ of all operators $\rho\in\T(\H_A)$ such that the limit in (\ref{Gen-d-rel}) exists w.r.t. to the trace norm. The set $\D(S)$ is the domain of the generator $S$.
\smallskip


Continuity of a quantum dynamical semigroup $\Phi_t$  w.r.t. the ECD norms means that\footnote{The definition of the ECD norm implies that  $\|\Phi_{s+t}-\Phi_{s}\|^G_{\diamond,E}\leq\|\Phi_t-\id_A\|^G_{\diamond,E}$ for any $t,s>0$.}
\begin{equation}\label{cont-rel}
 \lim_{t\rightarrow0^{+}} \|\Phi_t-\id_A\|^G_{\diamond,E}=0.
\end{equation}

\begin{property}\label{cont} \emph{Any quantum dynamical semigroup  $\{\Phi_t\}$ on $\T(\H_{A})$ is continuous w.r.t the ECD norm induced by a discrete positive operator $G$. If $\|S\|^G_{\diamond,E}<+\infty$ then}\footnote{This implies that for any separable Hilbert space $\H_R$ the superoperator $S\otimes \id_R$ is well defined on the set of  states $\rho$ in $\S(\H_{AR})$ such that $\Tr \rho_AG<+\infty$.}
\begin{equation}\label{cont-rel+}
\|\Phi_t-\id_A\|^G_{\diamond,E}\leq t \|S\|^G_{\diamond,E}\quad \forall t>0.
\end{equation}
\end{property}\smallskip

\begin{remark}\label{cont-r}
In contrast to the standard semigroup theory the condition $\|S\|^G_{\diamond,E}<+\infty$ is not necessary for continuity of $\{\Phi_t\}$ w.r.t. the ECD norm (see examples in Section 3). Estimates for $\|\Phi_t-\id_A\|^G_{\diamond,E}$ in the case $\|S\|^G_{\diamond,E}=+\infty$ can be obtained by the method used by Winter in \cite{W-EBN}. More detailed investigation of this problem had been made recently by Becker and Datta in \cite{Datta}.
\end{remark}\smallskip

\emph{Proof.} Since the convergence generated by any of the ECD norms coincides
with the strong convergence on subsets of Hermitian preserving superoperators bounded w.r.t. the diamond norm \cite[Proposition 3]{SCT},  any strongly continuous semigroup  $\{\Phi_t\}$ of quantum channels  is continuous w.r.t. the ECD norm.

If $\|S\|^G_{\diamond,E}<+\infty$  then Lemma \ref{L-1} below implies that
$$
\sup_{\rho\in \widehat{\S}^0_{G\otimes I_R, E}}\|\Phi_t\otimes \id_R(\rho)-\rho\|_1\leq t \|S\|^G_{\diamond,E},
$$
where $\widehat{\S}^0_{G\otimes I_R, E}$ is the set of all pure states in $\S(\H_{AR})$ having finite Schmidt rank such that $\Tr \rho_A G\leq E$.
Since the family $\{\Phi_t\}$ is bounded w.r.t. the diamond norm and the set $\S^0_{G\otimes I_R,E}$ is dense in the set
all pure states in $\S(\H_{AR})$  such that $\Tr \rho_AG\leq E$, the above relation and Lemma \ref{ecdn-def}B imply (\ref{cont-rel+}). $\square$\smallskip

\begin{lemma}\label{L-1} \emph{Let $\Phi_t$ be a quantum dynamical semigroup on $\T(\H_{A})$ such that the domain of its generator $S$ contains the set $\,\S_G$ of states $\rho$ with finite energy $\Tr \rho G$. Let $\,\H_R$ be a separable Hibert space. Then\smallskip
\begin{equation}\label{L-1-rel}
\Phi_t\otimes \id_R(\rho)-\rho=\int_0^t\Phi_s\otimes \id_R(S\otimes \id_R(\rho))ds
\end{equation}
for any pure state $\rho$ in $\S(\H_{AR})$ having finite Schmidt rank such that  $\Tr \rho_A G <+\infty$,\footnote{The integral in (\ref{L-1-rel}) and in all the formulae below is in the Bochner sense.} where
\begin{equation}\label{SI-def}
S\otimes \id_R(\rho)\doteq\sum_{i,j}S(|\varphi_i\rangle\langle\varphi_j|)\otimes |\psi_i\rangle\langle\psi_j|
\end{equation}
provided that $\,\rho=\sum_{i,j}|\varphi_i\rangle\langle\varphi_j|\otimes |\psi_i\rangle\langle\psi_j|$ is the
Schmidt representation of $\rho$.}
\end{lemma}\smallskip

\emph{Proof.} It suffices to note that the assumption $\S_G\subseteq\D(S)$ implies that $S\otimes \id_R(\rho)$ is well defined by formula (\ref{SI-def}) and that
\begin{equation}\label{lr-1}
\lim_{t\rightarrow0^{+}} \frac{\Phi_t\otimes \id_R(\rho)-\rho}{t}=S\otimes \id_R(\rho)
\end{equation}
for any pure state $\rho$ in $\S(\H_{AR})$ having finite Schmidt rank such that  $\Tr \rho_A G<+\infty$.  $\square$\smallskip

\begin{remark}\label{L-1-r}
Relation (\ref{L-1-rel}) holds for any $\rho\in\T(\H_{AR})$ for which (\ref{lr-1}) is valid.
\end{remark}
\medskip

Consider the question of differentiability of strongly continuous semigroups of quantum channels w.r.t. the ECD norm.\smallskip

Let $\Phi_t$ be a quantum dynamical semigroup on $\T(\H_{A})$ such that the domain of its generator $S$ contains the set $\,\S_G$ of states $\rho$ in $\S(\H_A)$ with finite energy $\Tr \rho G$. It means that
\begin{equation}\label{S-def}
  \lim\limits_{t\rightarrow0^{+}}\|(\Phi_t(\rho)-\rho)/t-S(\rho)\|_1=0\quad \textrm{for all} \;\, \rho\;\,\textrm{such that}\;\, \Tr \rho G<+\infty.
\end{equation}

We will say that the semigroup $\Phi_t$ is \emph{differentiable w.r.t. the ECD norm} if for a separable Hilbert space $\H_R$
the superoperator $S\otimes \id_R$ is well defined on the set of  states $\rho$ in $\S(\H_{AR})$ such that $\Tr \rho_A G<+\infty$
and
\begin{equation}\label{Dif-d}
  \lim\limits_{t\rightarrow0^{+}}\|(\Phi_t-\id_A)/t-S\|_{\diamond, E}^G=0\quad \textrm{for some}\quad E>0.
\end{equation}
This implies, in particular, that
\begin{equation*}
  \sup_{\rho\in\S(\H_A):\Tr \rho G\leq E}\|\Phi_t(\rho)-[\shs\rho+tS(\rho)]\|_1=o(t)\quad \textrm{as} \;\; t\rightarrow0^{+}.
\end{equation*}
It is clear that (\ref{Dif-d}) is substantially stronger than (\ref{S-def}). On the other hand,
the property (\ref{Dif-d}) is weaker than the differentiability of the semigroup $\Phi_t$ w.r.t. the diamond norm (which is equivalent to the boundedness of the generator $S$ and the representation $\Phi_t=e^{tS}$). So, the semigroups having property (\ref{Dif-d}) form a proper subclass of the class of all
strongly continuous semigroups which is \emph{substantially larger} than the subclass of uniformly continuous semigroups (see Corollary \ref{main-c} and Section 3).\smallskip

\begin{remark}\label{Dif-d-r} Proposition 3 in \cite{SCT}  does not imply that
(\ref{Dif-d}) follows from (\ref{S-def}), since the family of superoperators $\,(\Phi_t-\id_A)/t\,$ may not be bounded. This conclusion is confirmed by the examples considered in Section 3.
\end{remark}\smallskip

\begin{remark}\label{Dif-d-r+}
In contrast to the well known results concerning uniformly continuous semigroups, the condition $\|S\|^G_{\diamond,E}<+\infty$ is necessary but not
sufficient for differentiability of the semigroup $\Phi_t$ (see Section 3).
\end{remark}\smallskip

It turns out that necessary and sufficient conditions for the differentiability of a semigroup $\Phi_t$ w.r.t. the ECD norm can be expressed
in terms of the generator $S$ of this semigroup. For given separable infinite-dimensional Hibert space $\,\H_R$ and $E>0$ introduce the  sets
\begin{equation}\label{sp-set}
\widehat{\S}_{G\otimes I_R,E}\doteq\left\{\rho\in\S(\H_{AR})\,|\,\Tr \rho_AG\leq E, \rank\rho=1 \right\}
\end{equation}
and
\begin{equation}\label{sp-set+}
\T_{G\otimes I_R,E}\doteq\left\{\rho\in\T_{+}(\H_{AR})\,|\,\Tr \rho_AG\leq E, \Tr \rho\leq 1\right\}.
\end{equation}
\smallskip\pagebreak


\begin{theorem}\label{main} \emph{Let $\,\Phi_t$ be a quantum dynamical  semigroup on $\T(\H_{A})$ such that the domain of its generator $S$ contains all the states $\rho$ with finite energy $\Tr \rho G$. The following properties are equivalent:}
\begin{enumerate}[(i)]
   \item \emph{the semigroup $\Phi_t$ is differentiable w.r.t. the ECD norm, i.e. relation (\ref{Dif-d}) holds;}
   \item \emph{the superoperator $S\otimes \id_R$ is continuous on the set $\widehat{\S}_{G\otimes I_R,E}$ defined in (\ref{sp-set}) for some $E>0$;}
   \item \emph{the superoperator $S\otimes \id_R$ is uniformly continuous on the set $\T_{G\otimes I_R,E}$ defined in (\ref{sp-set+}) for any $E>0$ and  $\,\|S\|^G_{\diamond,E}=o(E)$ as $E\rightarrow+\infty$.}
 \end{enumerate}
\smallskip

\emph{Properties $(i)\textrm{-}(iii)$ hold provided that $\|S\|^G_{\diamond,E}=o(\sqrt{E})$ as $E\rightarrow+\infty$.}
\end{theorem}\medskip

\emph{Proof.} It is clear that (\ref{Dif-d}) implies that the superoperator $S$ belongs to the completion of the set of all Hermitian preserving completely bounded superoperators w.r.t. the ECD norm. So, the implication $\rm (i)\Rightarrow(iii)$  directly follows from the results of Section 5 in \cite{SPM}. The implication $\rm (iii)\Rightarrow(ii)$ is trivial.

To prove the implication $\rm (ii)\Rightarrow(i)$ consider the superoperator
$$
\Upsilon_t(\rho)=(1/t)\int_0^t(\Phi_s-\id_A)\otimes \id_R(S\otimes \id_R(\rho))ds
$$
well defined on the set $\widehat{\S}_{G\otimes I_R,E}$. Since
\begin{equation}\label{c-b}
 \|\Upsilon_t(\rho)-\Upsilon_t(\sigma)\|_1\leq 2\|S\otimes \id_R(\rho)-S\otimes \id_R(\sigma)\|_1,
\end{equation}
the superoperator $\Upsilon_t$ is continuous on the set $\widehat{\S}_{G\otimes I_R,E}$.

By Lemma \ref{L-1} we have
\begin{equation}\label{u-iden}
\Upsilon_t(\rho)=t^{-1}(\Phi_t\otimes \id_R(\rho)-\rho)-S\otimes \id_R(\rho)
\end{equation}
for any state $\rho$ in $\widehat{\S}_{G\otimes I_R,E}$ with finite Schmidt rank. Since the set of all such states is dense in
$\widehat{\S}_{G\otimes I_R,E}$, the continuity of $\Upsilon_t$ implies that (\ref{u-iden}) holds for all states in $\widehat{\S}_{G\otimes I_R,E}$.

Thus, by Lemma \ref{ecdn-def}B we have
\begin{equation}\label{u-rep}
\left\|\shs t^{-1}(\Phi_t-\id_{A})-S\shs\right\|^G_{\diamond,E}=
\sup_{\rho\in\widehat{\S}_{G,G_R,E}}\Upsilon_t(\rho),
\end{equation}
where $\widehat{\S}_{G,G_R,E}=\left\{\rho\in\S(\H_{AR})\,|\,\Tr \rho_A G\leq E, \Tr \rho_R G_R\leq E, \rank\rho=1 \right\}$ and $G_R$ is an operator
on $\H_R$ unitarily equivalent to $G$.

By the Lemma in \cite{H-c-w-c} and Corollary 6 in \cite{AQC} the set $\widehat{\S}_{G,G_R,E}$ is compact. Hence the superoperator $S\otimes \id_R$ is uniformly continuous on
$\widehat{\S}_{G,G_R,E}$ and the continuity bound  (\ref{c-b}) implies that
\begin{equation}\label{c-b+}
\|\Upsilon_t(\rho)-\Upsilon_t(\sigma)\|_1\leq f(\delta)
\end{equation}
for any $\rho$ and $\sigma$ in $\widehat{\S}_{G,G_R,E}$ such that $\|\rho-\sigma\|_1\leq \delta$ and all $t>0$, where $f(\delta)$ is a function vanishing as $\delta\rightarrow0^{+}$.

By using the compactness of $\widehat{\S}_{G,G_R,E}$ and the continuity bound (\ref{c-b+}) it is easy to show that the r.h.s. of
(\ref{u-rep}) tends to zero provided that $\Upsilon_t(\rho)$ tends to zero for any $\rho$ in $\widehat{\S}_{G,G_R,E}$.
The last property follows from the inequality
$$
\|\Upsilon_t(\rho)\|_1\leq(1/t)\int_0^t\|(\Phi_s-\id_A)\otimes \id_R(S\otimes \id_R(\rho))\|_1ds, \quad \rho\in\widehat{\S}_{G\otimes I_R,E},
$$
since $\|(\Phi_s-\id_A)\otimes \id_R(\sigma)\|_1$ tends to zero as $\,t\rightarrow0^{+}$ for any $\sigma\in \T(\H_{AR})$.\smallskip

If $\|S\|^G_{\diamond,E}=o(\sqrt{E})$ as $E\rightarrow+\infty$ then the superoperator $S\otimes \id_R$ is continuous on the set $\widehat{\S}_{G\otimes I_R,E}$ by Lemma \ref{S-cont} in Section 2.2. $\square$ \smallskip

It is well known that the generator of any uniformly continuous quantum dynamical semigroup has the standard (GKLS-) form
\begin{equation}\label{s-f}
S(\rho)=\sum_k V_k\rho V_k^*+K\rho+\rho K^*,
\end{equation}
where $V_k$ and $K$ are operators in $\B(\H_A)$ such that
\begin{equation}\label{ZT-cond}
  \sum_k \|V_k \varphi\|^2=-2\Re \langle\varphi|K\varphi\rangle
\end{equation}
for any $\varphi\in\H_A$ \cite{GKS,Lind}. Different generalizations of this representation to unbounded generators of quantum dynamical semigroups are considered in \cite{Ch,H-D,A&B,SHW}.\smallskip

The r.h.s. of  (\ref{s-f}) is well defined on the set $\S^0_G$ of finite rank states $\rho$ with finite energy $\Tr \rho G$ provided that all the operators $V_k$ and $K$ are defined on the domain of the operator $\sqrt{G}$ (it suffices to define the term $V_k\rho V_k^*$ by the formula similar to (\ref{ab-d+}) and to rewrite the term $\rho K^*$ as $[K\rho^*]^*$, cf.\cite{A&B,SHW}). So, in this case we may expect that the r.h.s. of  (\ref{s-f}) can be extended to a superoperator defined on the set $\S_G$ of all states with finite energy (which may be a generator of a strongly continuous semigroup differentiable w.r.t. the ECD norm).

Theorem \ref{main} implies the following\smallskip

\begin{corollary}\label{main-c} \emph{Let $\Phi_t$ be a quantum dynamical semigroup on $\T(\H_{A})$ with the generator $S$ defined by formula (\ref{s-f}) on the set $\,\S^0_G$, where $V_k$ and $K$ are  operators defined on $\D(\sqrt{G})$ and satisfying condition (\ref{ZT-cond}) for any $\varphi\in\D(\sqrt{G})$. Assume that the domain of the generator $\,S$ contains\footnote{It means that the condition (\ref{S-def}) holds.} the set $\,\S_G$ of all states with finite energy.}

A) \emph{If the operator $K$ is $\sqrt{G}$-infinitesimal   then the semigroup $\Phi_t$ is differentiable w.r.t. the ECD norm. 
In this case $\|S\|^G_{\diamond,E}\leq \|\{V_k\}\|_E^G+2\|K\|_E^G\leq 4\|K\|_E^G$, where}
\begin{equation}\label{K-c}
\|\{V_k\}\|_E^G\doteq\sup_{\rho\in\S(\H_A):\Tr \rho G\leq E}\sum_k \Tr V_k\rho V^*_k.
\end{equation}

B) \emph{If the operator $K$ is $\sqrt{G}$-bounded and the semigroup $\Phi_t$ is differentiable w.r.t. the ECD norm then the operator
$K$ and all the operators $V_k$ are $\sqrt{G}$-infinitesimal.}
\end{corollary}\smallskip

\emph{Proof.} B) Let $S_1(\rho)=\sum_k V_k\rho V_k^*$ and $S_2(\rho)=K\rho+[K\rho^*]^*$ be the positive and no-event parts of the
superoperator $S$ \cite{SHW}. If the operator $K$ is $\sqrt{G}$-bounded then condition (\ref{ZT-cond}) shows that
\begin{equation}\label{K-c}
\|\{V_k\}\|_E^G\leq 2\|K\|_E^G=o(E)\;\;\textup{ as }\;\;E\rightarrow+\infty,
\end{equation}
where the equality follows from the concavity of the function $E\mapsto[\|K\|_E^G]^2$. This implies, by formula (\ref{G-bound}), that all the operators $V_k$ are $\sqrt{G}$-infinitesimal. By Proposition 1 in \cite{SPM} relation (\ref{K-c}) shows that the superoperator $S_1$ belongs to the completion of the cone of completely positive superoperators w.r.t. the ECD norm. It follows that $S_1$ has property $\rm (iii)$ in Theorem \ref{main}. The differentiability of the semigroup $\Phi_t$  w.r.t. the ECD norm implies, by Theorem \ref{main}, that the same continuity property
holds for the superoperator $S_2=S-S_1$. Thus, Lemma \ref{disc} below shows that the operator $K$ is $\sqrt{G}$-infinitesimal.

A) If the operator $K$ is $\sqrt{G}$-infinitesimal then, by the above observation, all the operators $V_k$ are $\sqrt{G}$-infinitesimal as well. So, inequality (\ref{star-in}) with Lemma \ref{+I} and the inequality in (\ref{K-c}) imply that
$$
\|S\|^G_{\diamond,E}\leq \|\{V_k\}\|_E^G+2\|K\|_E^G\leq 4\|K\|_E^G=o\shs(\sqrt{E})\;\;\textup{ as }\;\;E\rightarrow+\infty,
$$
where the equality follows from (\ref{G-bound}). By  the last assertion of Theorem \ref{main} the semigroup $\Phi_t$  is differentiable w.r.t. the ECD norm. $\square$\smallskip

\begin{remark}\label{main-c-r}
If $K$ is not $\sqrt{G}$-bounded then it is easy to show that $\|S_2\|^G_{\diamond,E}$ can not be finite. So, if we assume
differentiability of the semigroup $\Phi_t$ in Corollary \ref{main-c} w.r.t. the ECD norm in this case, then $\|S_1\|^G_{\diamond,E}$ can not be finite
as well (since $\|S\|^G_{\diamond,E}$ must be finite). Formally, we can not exclude this possibility, but we have not managed to construct the corresponding examples.
\end{remark}\smallskip

\begin{lemma}\label{disc} \emph{Let $A$ be an operator defined on the set $\D(\sqrt{G})$ such that for any separable Hilbert
space $\H_R$ the operator $A\otimes I_R$ is well defined on the set $\D(\sqrt{G}\otimes I_R)$.}

\emph{If one of the superoperators $\Phi_{\pm}(\rho)=(A\otimes I_R) \rho\pm [(A\otimes I_R)\rho^*]^*$ is continuous
on the set $\T_{G\otimes I_R,E}$ defined in (\ref{sp-set+}) for some  $E>0$ then the operator $A$ is $\sqrt{G}$-infinitesimal.}
\end{lemma}\smallskip

\emph{Proof.}  Assume that the operator $G$ has the form (\ref{H-rep}) and $P_n=\sum_{k=0}^{n-1}|\tau_k\rangle\langle\tau_k|$
is  the projector on the subspace spanned by the vectors $\tau_0,..., \tau_{n-1}$.

If the operator $A$ does not lie in the space $\B^0_G(\H_A)$ of $\sqrt{G}$-infinitesimal operators then the sequence
$$
X_n=\sup\left\{\|(A\bar{P}_n)\otimes I_R\shs \varphi\|\,\left|\, \varphi\in\H_{AR}, \|\sqrt{G}\otimes I_R\shs\varphi\|^2\leq E, \|\varphi\|=1\right.\right\},
$$
where $\bar{P}_n=I_A-P_n$ does not tend to zero. Indeed, otherwise the sequence $\{AP_n\}$ of bounded operators  tends to the operator $A$ w.r.t. the norm $\|\cdot\|_E^G$
and hence $A$ belongs to the space $\B^0_G(\H_A)$ \cite[Remark 5]{ECN}. So, there is a sequence $\{\varphi_n\}$ of vectors in the unit ball of $\H_{AR}$ such that
$\|\sqrt{G}\otimes I_R\,\varphi_n\|^2\leq E$ for all $n$ and the sequence $\|(A\bar{P}_n)\otimes I_R\,\varphi_n\|$ does not tend to zero as $n\rightarrow\infty$. We may assume that the last sequence is bounded.\smallskip

Let $|\psi_n\rangle=\bar{P}_n\otimes I_R |\varphi_n\rangle$. Since $\|\sqrt{G}\otimes I_R\psi_n\|^2\leq E$ and $(\sqrt{G}\bar{P}_n)\otimes I_R \geq \sqrt{E_n}\bar{P}_n\otimes I_R $ for all $n$, we have $\|\psi_n\|\leq\sqrt{E/E_n}$. Hence, $\{\psi_n\}$ is a sequence tending to zero as $n\rightarrow\infty$. Let $\eta=\tau_0\otimes \upsilon$, where $\upsilon$ is a unit vector in $\H_R$.
We will assume that the vectors $|\alpha_n\rangle=A\otimes I_R |\psi_n\rangle$ do not converge to the vector $i|\eta\rangle$ (otherwise we can replace $\upsilon$). Consider the sequence of operators $\rho_n=\frac{1}{2}|\psi_n+\eta\rangle\langle\psi_n+\eta|\,$ in $\T_{G\otimes I_R,E}$ converging
to the operator $\rho_0=\frac{1}{2}|\eta\rangle\langle\eta|$. We have
$$
\begin{array}{c}
2A\otimes I_R\shs \rho_n+ 2[A\otimes I_R\shs \rho_n]^*=|\alpha_n+\beta\rangle\langle\psi_n+\eta|+|\psi_n+\eta\rangle\langle\alpha_n+\beta|=\\\\
|\alpha_n\rangle\langle\psi_n|+|\psi_n\rangle\langle\alpha_n|+
|\beta\rangle\langle\psi_n|+|\psi_n\rangle\langle\beta|+
\left[|\alpha_n\rangle\langle\eta|+|\eta\rangle\langle\alpha_n|\shs\right]+
|\beta\rangle\langle\eta|+|\eta\rangle\langle\beta|,
\end{array}
$$
where  $|\beta\rangle=A\otimes I_R |\eta\rangle$.
The first four  terms here tend to zero as $n\rightarrow\infty$, since $\{\psi_n\}$ tends to zero and the sequence
$\{\alpha_n\}$ is bounded. But the term in the square bracket does not tend to zero. So,
$\Phi_{+}(\rho_n)$ does not tend to
$$
\Phi_{+}(\rho_0)=\textstyle\frac{1}{2}[|\beta\rangle\langle\eta|+|\eta\rangle\langle\beta|].
$$
Similarly, one can prove the discontinuity of the superoperator $\Phi_{-}$ on the set $\T_{G\otimes I_R,E}$  if the operator $A$ is not  $\sqrt{G}$-infinitesimal. $\square$

\section{Higher order differentiability w.r.t. the ECD norm and the exponential representation}

In this section we obtain  necessary and sufficient conditions for $n$ order differentiability ($n\geq2$) of a (strongly continuous) quantum dynamical semigroups
 w.r.t. the metric induced by the ECD norm (\ref{ECDN-d}) assuming that $G$ is a discrete  operator (\ref{H-rep}) with $E_0=0$.

\smallskip

\begin{theorem}\label{main+} \emph{Let $\Phi_t$  be a quantum dynamical semigroup on  $\T(\H_{A})$ and $S$ the generator of this semigroup.
If $S^{k-1}(\rho)\in\D(S)$, $k=\overline{2,n}$, for any pure state $\rho$ with finite energy $\Tr \rho G$ and the superoperators $S^k\otimes \id_R$, $k=\overline{1,n}$, are continuous on the set $\,\widehat{\S}_{G\otimes I_R,E}$ defined in (\ref{sp-set})  for some $E>0$ then
\begin{equation}\label{main+r}
 \left\|\Phi_t-\left[\id_A+tS+\frac{t^2}{2}S^2+...+\frac{t^k}{k!}S^k\right]\right\|^G_{\diamond,E}\!=o\shs(t^k)\;\;\textrm{as}\;\; t\rightarrow0^{+},\;\; k=\overline{1,n},
 \end{equation}
for any $E>0$ and the l.h.s. of (\ref{main+r}) is bounded above by $\,2t^k\|S^k\|^G_{\diamond,E}/k!$.}\smallskip

\emph{If (\ref{main+r}) holds then the superoperators $S^k\otimes \id_R$, $k=1,2,...,n$, are uniformly continuous on the set $\T_{G\otimes I_R,E}$ defined in (\ref{sp-set+}) and  $\|S^k\|^G_{\diamond,E}=o(E)$, $k=\overline{1,n}$, as $E\rightarrow+\infty$.}
\smallskip
\end{theorem}

\emph{Proof.} By sequentially applying Lemma \ref{L-3} below one can obtain
$$
\widehat{\Phi}_t(\rho)=\rho+t\widehat{S}(\rho)+\frac{t^2}{2}\widehat{S}^2(\rho)+..+\frac{t^{k-1}}{(k-1)!}\widehat{S}^{k-1}(\rho)
+\int_0^t\widehat{\Phi}_s(\widehat{S}^k(\rho))\frac{(t-s)^{k-1}}{(k-1)!}\,ds,\;\; k=\overline{1,n},
$$
for any state $\rho$ in $\widehat{\S}_{G\otimes I_R,E}$, where $\widehat{\Phi}_t$ and $\widehat{S}^k$ denote, respectively, the superoperators $\Phi_t\otimes \id_R$ and $S^k\otimes \id_R$. By using  the superoperator
$$
\Gamma_{t,k}(\rho)=\widehat{\Phi}_t(\rho)-\rho-t\widehat{S}(\rho)-\frac{t^2}{2}\widehat{S}^2(\rho)-..-\frac{t^{k}}{k!}\widehat{S}^{k}(\rho)
$$
well defined on the set $\widehat{\S}_{G\otimes I_R,E}$ the above relation can be written as
$$
\Gamma_{t,k}(\rho)=\int_0^t(\Phi_s-\id_A)\otimes \id_R(S^k\otimes \id_R(\rho))\frac{(t-s)^{k-1}}{(k-1)!}\,ds,\quad k=\overline{1,n},
$$
It follows that for any states $\rho$ and $\sigma$ in $\widehat{\S}_{G\otimes I_R,E}$ we have
\begin{equation}\label{c-b++}
 t^{-k}\|\Gamma_{t,k}(\rho)-\Gamma_{t,k}(\sigma)\|_1\leq (2/k!)\|S^k\otimes \id_R(\rho)-S^k\otimes \id_R(\sigma)\|_1.
\end{equation}

Lemma \ref{ecdn-def}B implies that
\begin{equation}\label{u-rep+}
\left\|\Phi_t-\left[\id_A+tS+\frac{t^2}{2}S^2+...+\frac{t^k}{k!}S^k\right]\right\|^G_{\diamond,E}=
\sup_{\rho\in\widehat{\S}_{G,G_R,E}}\Gamma_{t,k}(\rho),
\end{equation}
where $\widehat{\S}_{G,G_R,E}=\left\{\rho\in\S(\H_{AR})\,|\,\Tr \rho_AG\leq E, \Tr \rho_R G_R\leq E, \rank\rho=1 \right\}$ and $G_R$ is an operator
on $\H_R$ unitarily equivalent to $G$.

By the Lemma in \cite{H-c-w-c} and Corollary 6 in \cite{AQC} the set $\widehat{\S}_{G,G_R,E}$ is compact. Hence the superoperator $S^k\otimes \id_R$ is uniformly continuous on this set and continuity bound  (\ref{c-b++}) implies that
\begin{equation}\label{c-b+++}
t^{-k}\|\Gamma_{t,k}(\rho)-\Gamma_{t,k}(\sigma)\|_1\leq g_k(\delta)
\end{equation}
for any $\rho$ and $\sigma$ in $\widehat{\S}_{G,G_R,E}$ such that $\|\rho-\sigma\|_1\leq \delta$ and all $t>0$, where $g_k(\delta)$ is a function vanishing as $\delta\rightarrow0^+$.

The compactness of $\widehat{\S}_{G,G_R,E}$ and continuity bound (\ref{c-b+++}) imply that  the r.h.s. of
(\ref{u-rep+}) is $o(t^k)$ as $t\rightarrow0^+$ if and only if $\,t^{-k}\|\Gamma_{t,k}(\rho)\|_1$ tends to zero  as $t\rightarrow0^+$ for any $\rho$ in $\widehat{\S}_{G,G_R,E}$. The last property can be easily proved by using the inequality
$$
\|\Gamma_{t,k}(\rho)\|_1\leq\int_0^t\|(\Phi_t-\id_A)\otimes \id_R(S^k\otimes \id_R(\rho))\|_1\frac{(t-s)^{k-1}}{(k-1)!}\,ds,
$$
since $\|(\Phi_s-\id_A)\otimes \id_R(\sigma)\|_1$ tends to zero as $\,t\rightarrow0^{+}$ for any $\sigma\in \T(\H_{AR})$.\smallskip

By this inequality $\|\Gamma_{t,k}(\rho)\|_1\leq (2t^k/k!)\|S^k\otimes \id_R(\rho)\|_1$ for any state $\rho$ in $\widehat{\S}_{G\otimes I_R,E}$. So,
it follows from (\ref{u-rep+}) that the l.h.s. of (\ref{main+r}) is bounded above by $2t^k\|S^k\|^G_{\diamond,E}/k!$.\smallskip

If (\ref{main+r}) holds then it is easy to show iteratively  that the superoperators $S^k$, $k=\overline{1,n}$, belong to the completion of the set of all Hermitian preserving completely bounded superoperators w.r.t. the ECD norm. So, the last assertion of the theorem directly follows from the results of Section 5 in \cite{SPM}. $\square$ \smallskip

\begin{lemma}\label{L-3} \emph{If the assumptions of Theorem \ref{main+} hold then
\begin{equation}\label{L-3-r}
\Phi_t\otimes \id_R(S^{k-1}\otimes \id_R(\rho))-S^{k-1}\otimes \id_R(\rho)=\int_0^t\Phi_s\otimes \id_R(S^k\otimes \id_R(\rho))ds
\end{equation}
for  $k=\overline{1,n}$ and any state $\rho$ in $\,\widehat{\S}_{G\otimes I_R,E}$.}
\end{lemma} \smallskip

\emph{Proof.} Let $\rho$ be a state in $\widehat{\S}_{G\otimes I_R,E}$ with finite Schmidt rank $n$. It can be represented as
$$
\rho=\sum_{i,j=1}^n |\varphi_i\rangle\langle\varphi_j|\otimes |\psi_i\rangle\langle\psi_j|,
$$
where $\{\varphi_i\}$ and $\{\psi_i\}$ are orthogonal sets of vectors in $\D(\sqrt{G})$ and $\H_R$ correspondingly. Then
$$
S^{k-1}\otimes \id_R(\rho)=\sum_{i,j=1}^n S^{k-1}(|\varphi_i\rangle\langle\varphi_j|)\otimes |\psi_i\rangle\langle\psi_j|.
$$
Since $S^{k-1}(|\varphi_i\rangle\langle\varphi_j|)\in\D(S)$ for all $i$ and $j$ by the assumption, it follows that
$$
\lim_{t\rightarrow0^+}(1/t)(\Phi_t\otimes \id_R(S^{k-1}\otimes \id_R(\rho))-S^{k-1}\otimes \id_R(\rho))=S^{k}\otimes \id_R(\rho).
$$
So, by Remark \ref{L-1-r}, Lemma \ref{L-1} implies that (\ref{L-3-r}) holds for the state $\rho$.  Since the set of states in $\widehat{\S}_{G\otimes I_R,E}$ with finite Schmidt rank is dense in  $\widehat{\S}_{G\otimes I_R,E}$, the validity of equality (\ref{L-3-r}) for any state in $\widehat{\S}_{G\otimes I_R,E}$ follows from the continuity of both sides of this equality on $\widehat{\S}_{G\otimes I_R,E}$. $\square$\smallskip

\begin{corollary}\label{main+c} \emph{Let $\,\Phi_t$ be a quantum dynamical semigroup on $\T(\H_{A})$ with the generator $S$ having the properties:
\begin{itemize}
  \item $S^{n-1}(\rho)\in\D(S)$, $n\in\mathbb{N}$, for any pure state $\rho$ with finite energy $\Tr \rho G$;
  \item the superoperators $S^n\otimes \id_R$, $n\in\mathbb{N}$, are continuous on the set $\,\widehat{\S}_{G\otimes I_R,E}$ defined in (\ref{sp-set}) for some $E>0$;
  \item $\|S^n\|^G_{\diamond,E}/n!\,$ tends to zero as $\,n\rightarrow+\infty\,$  for some $\,E>0$.
\end{itemize}
Then
\begin{equation}\label{main+c-r}
 \Phi_t|_{\S_{G}}=e^{tS}|_{\S_{G}},\quad t>0,
 \end{equation}
where $\Lambda|_{\S_{G}}$ is the restriction of $\,\Lambda$ to the set $\,\S_{G}\doteq \{\rho\in\S(\H_A)\,|\,\Tr \rho G<+\infty\}$ and
$e^{tS}$ denotes the series
$$
\id_A+tS+\frac{t^2}{2}S^2+...+\frac{t^k}{k!}S^k+...
$$
converging w.r.t. the norm $\|\cdot\|^G_{\diamond,E}$.}
\end{corollary}

\section{Examples}

\subsection{The unitary group $e^{-iAt}$}

Consider the group of unitary channels $\Lambda^A_t(\rho)=e^{-iAt}\rho\, e^{iAt}$, where $A$ is a self-adjoint operator on $\H$.
To specify the generator of this semigroup and its domain we will need the following \smallskip

\begin{definition}\label{s-s-o} \cite[II.8]{aspecty} A densely defined operator $A$ is called \emph{square-summable} w.r.t. a state $\rho$ if
the operator $\sqrt{\rho}A$ is (extended to) a Hilbert-Schmidt operator.
\end{definition}\smallskip

It is easy to see that an operator $A$ is \emph{square-summable} w.r.t. a state $\rho$ if and only if the operator $A\rho A$ is (extended to)
a trace-class operator. The last property is equivalent to the finiteness of $\,\Tr \rho A^2\doteq \sup_n \Tr \rho P_nA^2$, where $P_n$ is the spectral projector of $A$ corresponding to the interval $[0,n]$. \smallskip

\begin{lemma}\label{l-ex-1} \cite[Proposition VI.3.1]{aspecty} \emph{The domain of the generator $S_{A}$ of the semigroup $\Lambda^A_t$ contains all the states $\rho$ such that the operator $A$ is square-summable w.r.t. the state $\rho
$. The action of $S_{A}$  on such a state $\rho $ is given by}
\begin{equation}\label{s-a}
S_{A}(\rho )=i\left((\sqrt{\rho}A)^{\ast}\!\sqrt{\rho}-\sqrt{\rho}(\sqrt{\rho}A)\right).
\end{equation}
\end{lemma}

The operator on the right-hand side of (\ref{s-a}) is apparently
trace-class, giving a well-defined version for the expression
\begin{equation*}
i(A\rho-\rho A)=i[A,\rho],
\end{equation*}
which holds literally in the case of bounded operator $A$.

Lemma \ref{l-ex-1} and Lemma \ref{Gb} in Section 2.3 show that $\D(S_A)$ contains all states $\rho$ with finite energy $\Tr\rho G$ if and only if
the operator $A$ is $\sqrt{G}$-bounded. By using the inequality (\ref{star-in}) with Lemma \ref{+I} and Remark \ref{qsl-r} it is easy to show that in this case $\|S_A\|_{\diamond,E}^G\leq 2\|A\|^G_E$. So, the results of the previous sections  imply the following \smallskip

\begin{property}\label{ex-1} \emph{Let $G$ be a positive discrete unbounded operator (\ref{H-rep}) with $E_0=0$.}\smallskip

A) \emph{The group  $\{\Lambda^A_t\}$ is  continuous w.r.t. the ECD norm induced by  $G$. If the operator $A$ is $\sqrt{G}$-bounded then}\footnote{If $A=G$ then the r.h.s. of (\ref{n-cont-rel+}) is equal to $+\infty$. The estimate for $\|\Lambda^G_t-\id\|^G_{\diamond,E}$ is obtained in \cite{W-EBN}. It is refined in \cite{Datta}, where the estimates for $\|\Lambda^G_t-\id\|^{G^{2\alpha}}_{\diamond,E^{2\alpha}}$, $\alpha\in(0,1]$,  are also obtained. If  $A=\sqrt{G}$ then (\ref{n-cont-rel+}) coincides with the estimate in Proposition 3.2 in \cite{Datta} for $\alpha=1$ (since $\|\sqrt{G}\|_E^G=\sqrt{E}$).}
\begin{equation}\label{n-cont-rel+}
\|\Lambda^A_t-\id\|^G_{\diamond,E}\leq 2t\|A\|_E^G\quad \forall t>0.
\end{equation}

B) \emph{The group  $\{\Lambda^A_t\}$ is differentiable w.r.t. the ECD norm induced by  $G$ if and only if $A$ is a $\sqrt{G}$-infinitesimal operator.}\smallskip

C) \emph{If $A^{n}$ is a $\sqrt{G}$-infinitesimal operator for some $\,n\in\mathbb{N}\,$ then
\begin{equation}\label{ex-1-rel}
 \left\|\Lambda^A_t-\left[\id+tS_A+\frac{t^2}{2}S_A^2+...+\frac{t^k}{k!}S_A^k\right]\right\|^G_{\diamond,E}\!=o\shs(t^k)\;\;\textrm{as}\;\; t\rightarrow0^{+},\;\; k=\overline{1,n};
 \end{equation}
for any $E>0$ and the l.h.s. of (\ref{ex-1-rel}) is bounded above by $\,2(2t)^k\|A^{k}\|^G_E/k!$.}\smallskip

D) \emph{If $A^n$ is a $\sqrt{G}$-bounded operator for any $\,n\in\mathbb{N}\,$ and $\,\|A^n\|^G_E\leq C[n!]^p$ for some $\,E,C>0\,$ and $\,p<1\,$ then
\begin{equation}\label{ex-1-exp}
 \Lambda^A_t|_{\S_{G}}=e^{tS_A}|_{\S_{G}}\quad\textit{for any }\;\;t>0,
 \end{equation}
i.e. the restriction of the group $\Lambda^A_t$ to the set $\,\S_{G}\doteq \{\rho\in\S(\H)\,|\,\Tr \rho G<+\infty\}$ is represented by the series
$$
\id+tS_A+\frac{t^2}{2}S_A^2+...+\frac{t^k}{k!}S_A^k+...
$$
converging w.r.t. the norm $\|\cdot\|^G_{\diamond,E}$.}
\end{property} \medskip

Proposition \ref{ex-1} shows that:
\begin{itemize}
  \item the group $\Lambda^{G^\alpha}_t(\rho)=e^{-iG^\alpha t}\rho\, e^{iG^\alpha t}$ is  differentiable   w.r.t the ECD norm induced by a positive operator $G$ \textbf{if and only if} $\,\alpha<1/2$;
  \item if $\,\alpha<1/2^n$ then relations (\ref{ex-1-rel}) hold for the group $\Lambda^{G^\alpha}_t$.
\end{itemize}
\smallskip

To construct a group $\Lambda^{A}_t$ having the exponential representation (\ref{ex-1-exp}) take $A=\sqrt{\ln G}$. Since $\ln^n G\leq n!G$, we have
$$
\|[\sqrt{\ln G}]^n\|^G_E\leq \sqrt{n!}\|\sqrt{G}\|^G_E=\sqrt{n!E}<+\infty\quad \forall n.
$$
Hence in this case formula (\ref{G-bound}) and Lemma \ref{kln} imply that $\,A^n$ is a $\sqrt{G}$-infinitesimal operator for any $n\in\mathbb{N}$. Thus, all the conditions of Proposition \ref{ex-1}D hold. \smallskip

\emph{Proof of Proposition \ref{ex-1}.} A) This assertion directly follows from Proposition  \ref{cont} and the above upper bound on  $\|S_A\|_{\diamond,E}^G$. \smallskip

B) This assertion follows from Corollary \ref{main-c} and the remark before the proposition.\smallskip

C) If $A^{n}$ is a $\sqrt{G}$-infinitesimal operator then $\varphi\in\D(A^{n})$ for any pure state $\,\rho=|\varphi\rangle\langle\varphi|\,$ with finite energy $\Tr\rho G=\|\sqrt{G}\varphi\|^2$. Since
$$
S_A(|\psi\rangle\langle\psi|)=\frac{1}{2}\shs(B|\psi\rangle\langle\psi| B^*-B^*|\psi\rangle\langle\psi| B),\qquad B=A-iI,
$$
for any vector $\psi\in\D(A)$, it is easy to show that for any pure state $\,\rho\,$ with finite energy we have $S_A^{k-1}(\rho)=X_{k,+}^{\rho}-X_{k,-}^{\rho}$, where $X^{\rho}_{k,+}$ and
$X^{\rho}_{k,-}$  are operators in $\T_+(\H)$ such that $\Tr X^{\rho}_{k,\pm}A^2<+\infty$, $k=\overline{2,n}$.
So, Lemma \ref{l-ex-1} implies that $S_A^{k-1}(\rho)\in\D(S_A)$, $k=\overline{2,n}$ for any such state $\,\rho$. By using inequality (\ref{star-in}) with Lemma \ref{+I} and Lemma \ref{kln} we obtain
\begin{equation}\label{S-up}
 \|S^k_A\|_{\diamond,E}^G\leq \sum\limits_{i=0}^k \binom{i}{k}\|A^i\|^G_E \|A^{k-i}\|^G_E\leq 2^k\|A^{k}\|^G_E=o\shs(\sqrt{E})\quad \textrm{as}\quad E\rightarrow+\infty,
\end{equation}
where the last equality follows from (\ref{G-bound}), since $A^k$ is a $\sqrt{G}$-infinitesimal operator for each $k=\overline{1,n}$ by Lemma \ref{kln}.  By Lemma \ref{S-cont} the superoperators $S_A^k\otimes \id_R$,
$k=\overline{1,n}$, are continuous on the set $\widehat{\S}_{G\otimes I_R,E}$ defined in (\ref{sp-set}).
Thus, this assertion follows from Theorem \ref{main+}. \smallskip

D) This assertion follows from  Corollary \ref{main+c} and Lemma \ref{kln}, since the condition $\,\|A^n\|^G_E\leq C[n!]^p$ and estimate (\ref{S-up}) imply that  $\|S_A^n\|^G_{\diamond,E}/n!$ tends to zero as $n\rightarrow+\infty$  for any $E>0$. $\square$

\subsection{The Gaussian convolutional semigroup}

Let  $A$ be a self-adjoint operator on $\H$. Consider the semigroup of quantum  channels
\begin{equation}\label{gaus}
\Xi^A_t(\rho)=\frac{1}{\sqrt{2\pi t}}\int_{-\infty}^{+\infty} e^{-iAx}\rho\, e^{iAx} e^{-\frac{x^2}{2t}} dx.
\end{equation}

We will use the following lemma proved in the Appendix.\smallskip

\begin{lemma}\label{l-ex-2} \emph{The domain of the generator $Z_{A}$ of the semigroup $\Xi^A_t$ contains all the states
$\rho $ such that the operator $A^{2}$ is square-summable w.r.t. the state $\rho
$.\footnote{See Definition \ref{s-s-o} in the previous subsection.} The action of $Z_{A}$  on such
a state $\rho $ is given by}
\begin{equation}
Z_{A}(\rho )=(\sqrt{\rho }A)^{\ast }\sqrt{\rho }A-\frac{1}{2}\left[ (\sqrt{
\rho }A^{2})^{\ast }\sqrt{\rho }+\sqrt{\rho }(\sqrt{\rho }A^{2})\right] .
\label{gen}
\end{equation}
\end{lemma}

The operator on the right-hand side of (\ref{gen}) is apparently
trace-class, giving a well-defined version for the expression
\begin{equation*}
A\rho A-\frac{1}{2}\left(A^2\rho+\rho A^2\right)=\frac{1}{2}\,[A,[\rho, A]],
\end{equation*}
which holds literally in the case of bounded operator $A$.

Lemma \ref{l-ex-2} and Lemma \ref{Gb} in Section 2.3 show that $\D(Z_A)$ contains all states $\rho$ with finite energy $\Tr\rho G$ if and only if
the operator $A^2$ is $\sqrt{G}$-bounded. By using inequality (\ref{star-in}) with Lemma \ref{+I} and Remark \ref{qsl-r} it is easy to show that in this case
$$
\|Z_A\|_{\diamond,E}^G\leq [\|A\|^G_E]^2+\|A^2\|^G_E\leq 2\|A^2\|^G_E.
$$
So, the results of the previous sections  imply the following \smallskip

\begin{property}\label{ex-2} \emph{Let $G$ be a positive discrete unbounded operator (\ref{H-rep}) with $E_0=0$.}\smallskip

A) \emph{The semigroup  $\{\Xi^A_t\}$ is  continuous w.r.t. the ECD norm induced by  $G$. If the operator $A^2$ is $\sqrt{G}$-bounded  then}\footnote{Estimates for $\|\Xi^A_t-\id\|^G_{\diamond,E}$ in the case $\|A^2\|_E^G\leq+\infty$ can be obtained by using \cite[Theorem 1]{Datta}.}
\begin{equation}\label{cont-rel++}
\|\Xi^A_t-\id\|^G_{\diamond,E}\leq t\left[[\|A\|^G_E]^2+\|A^2\|^G_E\right]\quad \forall t>0.
\end{equation}

B) \emph{The semigroup  $\{\Xi^A_t\}$ is differentiable w.r.t. the ECD norm induced by  $G$ if and only if $A^2$ is a $\sqrt{G}$-infinitesimal operator.}\smallskip

C) \emph{If $A^{2n}$ is $\sqrt{G}$-infinitesimal operator for some $n\in\mathbb{N}$ then
\begin{equation}\label{ex-2-rel}
 \left\|\Xi^A_t-\left[\id+tZ_A+\frac{t^2}{2}Z_A^2+...+\frac{t^k}{k!}Z_A^k\right]\right\|^G_{\diamond,E}\!=o\shs(t^k)\;\;\textrm{as}\;\; t\rightarrow0^{+},\;\; k=\overline{1,n};
 \end{equation}
for any $E>0$ and the l.h.s. of (\ref{ex-2-rel}) is bounded above by $\,2(2t)^k\|A^{2k}\|^G_E/k!$.}\smallskip

D)\emph{ If $\,A^n$ is $\sqrt{G}$-bounded operator for any $n\in\mathbb{N}$ and $\,\|A^{2n}\|^G_E\leq C[n!]^p$ for some $\,E,C>0\,$ and $\,p<1\,$ then
\begin{equation}\label{ex-2-exp}
 \Xi^A_t|_{\S_{G}}=e^{tZ_A}|_{\S_{G}}\quad\textit{for any }\; \;t>0,
 \end{equation}
i.e. the restriction of the semigroup $\Xi^A_t$ to the set $\,\S_{G}\doteq \{\rho\in\S(\H)\,|\,\Tr \rho G<+\infty\}$ is represented by the series
$$
\id+tZ_A+\frac{t^2}{2}Z_A^2+...+\frac{t^k}{k!}Z_A^k+...
$$
converging w.r.t. the norm $\|\cdot\|^G_{\diamond,E}$.}
\end{property} \medskip

Proposition \ref{ex-2} shows that:
\begin{itemize}
  \item the semigroup $\Xi^{G^\alpha}_t$ is  differentiable   w.r.t. the ECD norm indiced by a positive operator $G$ \textbf{if and only if} $\,\alpha<1/4$;
  \item if $\,\alpha<1/4^n$ then relations (\ref{ex-2-rel}) hold for the semigroup $\Xi^{G^\alpha}_t$.
\end{itemize}
\smallskip

By repeating the arguments after Proposition \ref{ex-1} one can show that the semigroup $\Xi^{A}_t$
satisfies the condition of Proposition \ref{ex-2}D if $A=\sqrt[4]{\ln G}$. So, in this case  the exponential
representation (\ref{ex-2-exp}) is valid.\smallskip

\emph{Proof of Proposition \ref{ex-2}.} A) This assertion directly follows from Proposition  \ref{cont} and the above upper bound on  $\|Z_A\|_{\diamond,E}^G$. \smallskip

B) This assertion follows from Corollary \ref{main-c} and the remark before the proposition.\smallskip

C) Note first that $Z_A(\rho)=S^2_A(\rho)/2$ and hence $Z^n_A(\rho)=S^{2n}_A(\rho)/2^n$, where $S_A$ is the superoperator defined in (\ref{s-a}).

If $A^{2n}$ is a $\sqrt{G}$-infinitesimal operator then, by using the arguments from the proof of Proposition 2C, it is easy to show that for any pure state $\,\rho\,$ with finite energy $\,\Tr\rho G$ we have
$Z_A^{k-1}(\rho)=2^{1-k}S_A^{2(k-1)}(\rho)=Y_{k,+}^{\rho}-Y_{k,-}^{\rho}$, where $Y^{\rho}_{k,+}$ and
$Y^{\rho}_{k,-}$  are operators in $\T_+(\H)$ such that $\Tr Y^{\rho}_{k,\pm}A^4<+\infty$, $k=\overline{2,n}$.
So, Lemma \ref{l-ex-2} implies that $Z_A^{k-1}(\rho)\in\D(Z_A)$, $k=\overline{2,n}$, for any such state $\rho$.  By using inequality (\ref{star-in}) with Lemma \ref{+I} and Lemma \ref{kln} we obtain
\begin{equation}\label{Z-up}
\begin{array}{rl}
 \|Z^k_A\|_{\diamond,E}^G=2^{-k}\|S^{2k}_A\|_{\diamond,E}^G\!\! & \leq \,2^{-k}\sum\limits_{i=0}^{2k}\binom{i}{2k}\|A^i\|^G_E \|A^{2k-i}\|^G_E
 \\\\ & \leq \,2^{k}\|A^{2k}\|^G_E=o\shs(\sqrt{E})\quad \textrm{as}\quad E\rightarrow+\infty,
\end{array}
\end{equation}
where the last equality follows from (\ref{G-bound}), since $A^{2k}$ is a $\sqrt{G}$-infinitesimal operator for each $k=\overline{1,n}$ by Lemma \ref{kln}.  By Lemma \ref{S-cont} the superoperators $Z_A^k\otimes \id_R$,
$k=\overline{1,n}$, are continuous on the set $\widehat{\S}_{G\otimes I_R,E}$ defined in (\ref{sp-set}).
Thus, this assertion follows from Theorem \ref{main+}. \smallskip

D) This assertion follows from  Corollary \ref{main+c} and Lemma \ref{kln}, since the condition $\,\|A^{2n}\|^G_E\leq C [n!]^p$ and the estimate (\ref{Z-up}) imply that  $\|Z_A^n\|^G_{\diamond,E}/n!$ tends to zero as $n\rightarrow+\infty$  for any $E>0$. $\square$

\section*{Appendix: Proof of Lemma \ref{l-ex-2}}

By making change of variable $\,u=x/\sqrt{t}\,$ in the integral (\ref{gaus}), we obtain
\begin{equation}\label{inta}
t^{-1}[\,\Xi _{t}^{A}(\rho )-\rho\,]=\frac{1}{\sqrt{2\pi }}\int_{-\infty
}^{+\infty }t^{-1}\left[ e^{-i\sqrt{t}uA}\rho \,e^{i\sqrt{t}uA}-\rho \right]
e^{-\frac{u^{2}}{2}}du.
\end{equation}
Expanding the exponent, we have
\begin{equation}
e^{i\sqrt{t}uA}=I+i\sqrt{t}uA-\frac{1}{2}tu^{2}A^{2}+tu^{2}A^{2}F(\sqrt{t}
uA),  \label{expa}
\end{equation}
where the function
\begin{equation*}
F(u)=u^{-2}\left( e^{iu}-1-iu+\frac{1}{2}u^{2}\right)
\end{equation*}
is uniformly bounded and $F(u)\rightarrow 0$ as $u\rightarrow 0$. It follows
that operator-valued function $F(\sqrt{t}uA)$ is uniformly bounded in the
operator norm and $F(\sqrt{t}uA)\rightarrow 0$ strongly as $t\rightarrow 0$
for any fixed $u$ (cf. the proof of Proposition VI.3.1 in \cite{aspecty}) .

By inserting the expansion (\ref{expa}) into (\ref{inta}) we obtain
\begin{equation*}
\frac{1}{\sqrt{2\pi }}\int_{-\infty }^{+\infty }t^{-1}\left[ \left(
h.c.\right) \left( \sqrt{\rho }+i\sqrt{t}u\sqrt{\rho }A-\frac{1}{2}tu^{2}
\sqrt{\rho }A^{2}+tu^{2}\sqrt{\rho }A^{2}F(\sqrt{t}uA)\right) -\rho \right]
e^{-\frac{u^{2}}{2}}du,
\end{equation*}
where $\left( h.c.\right) $ is the Hermitian conjugate to the subsequent
expression in the round bracket. Let $\sigma _{1}=\sqrt{\rho }
A$ and $\,\sigma _{2}=\sqrt{\rho }A^{2}$ be Hilbert-Schmidt operators. Taking
into account that the odd moments of the standard normal distribution
vanish, second moment is 1 and fourth moment is 3, we obtain
\begin{eqnarray*}
&&Z_{A}(\rho )+\frac{3t}{4}\,\sigma _{2}^{\ast }\,\sigma _{2}+\frac{t}{\sqrt{
2\pi }}\int_{-\infty }^{+\infty }\left( \,\overline{F}(\sqrt{t}uA)\sigma
_{2}^{\ast }\sigma _{2}F(\sqrt{t}uA)\right) u^{4}e^{-\frac{u^{2}}{2}}du \\
&&\frac{1}{\sqrt{2\pi }}\int_{-\infty }^{+\infty }\left( \,\overline{F}(
\sqrt{t}uA)\sigma _{2}^{\ast }\right) \left( \sqrt{\rho }+i\sqrt{t}u\sigma
_{1}-\frac{1}{2}tu^{2}\sigma _{2}\right) u^{2}e^{-\frac{u^{2}}{2}}du+h.c.,
\end{eqnarray*}
where now $h.c.$ denotes the Hermitian conjugate of the second integral.
Taking into account the uniform boundedness of $\left\Vert F(\sqrt{t}
uA)\right\Vert ,$ we conclude that the limit of this whole expression in the
trace norm as $t\rightarrow 0$ is
\begin{equation*}
Z_{A}(\rho )+\lim_{t\rightarrow 0}\left[ \frac{1}{\sqrt{2\pi }}\int_{-\infty
}^{+\infty }\,\overline{F}(\sqrt{t}uA)\sigma _{2}^{\ast }\sqrt{\rho }
u^{2}e^{-\frac{u^{2}}{2}}du+h.c.\right] .
\end{equation*}
But $\left\Vert \overline{F}(\sqrt{t}uA)\sigma _{2}^{\ast }\sqrt{\rho }
\right\Vert _{1}\leq \left\Vert \overline{F}(\sqrt{t}uA)\right\Vert \left\Vert \sigma _{2}^{\ast }\sqrt{\rho }
\right\Vert _{1}$ is uniformly bounded and
$$\lim_{t\rightarrow 0}\left\Vert
\overline{F}(\sqrt{t}uA)\sigma _{2}^{\ast }\sqrt{\rho }\right\Vert _{1}=0$$
for each $u$ (this follows from the strong convergence $F(\sqrt{t}
uA)\rightarrow 0).$ So,  by the dominated convergence theorem the limit of
the integral in the trace norm is equal to zero. Thus
\begin{equation*}
\lim_{t\rightarrow 0}t^{-1}[\Xi _{t}^{A}(\rho )-\rho ]=Z_{A}(\rho )
\end{equation*}
in the  trace norm for any state $\rho $ satisfying the condition of Lemma \ref{l-ex-2}.
\bigskip

The authors are grateful to A.M.Chebotarev for useful communication.

\end{document}